\documentclass{article}

\usepackage{arxiv}

\usepackage[utf8]{inputenc} 
\usepackage[T1]{fontenc}    
\usepackage{hyperref}       
\usepackage{url}            
\usepackage{booktabs}       
\usepackage{amsfonts}       
\usepackage{nicefrac}       
\usepackage{microtype}      
\usepackage{lipsum}		
\usepackage{xcolor}
\usepackage{amsmath}
\usepackage{bm} 
\usepackage[colorinlistoftodos,prependcaption,textsize=scriptsize]{todonotes}

\title{BioImageLoader: easy handling of bioimage datasets for machine learning}

\date{} 					

\author{
  Seongbin Lim\\
  Laboratoire d'Optique et Biosciences\\
  CNRS, INSERM,
  École Polytechnique, Institut Polytechnique de Paris\\
  91120 Palaiseau, France\\
  \texttt{sungbin246@gmail.com}\\
    \And
  Xingjian Zhang\\
  Laboratoire d'Optique et Biosciences\\
  CNRS, INSERM,
  École Polytechnique, Institut Polytechnique de Paris\\
  91120 Palaiseau, France\\
    \And
  Emmanuel Beaurepaire\\
  Laboratoire d'Optique et Biosciences\\
  CNRS, INSERM,
  École Polytechnique, Institut Polytechnique de Paris\\
  91120 Palaiseau, France\\
    \And
  Anatole Chessel\\
  Laboratoire d'Optique et Biosciences\\
  CNRS, INSERM,
  École Polytechnique, Institut Polytechnique de Paris\\
  91120 Palaiseau, France\\
  \texttt{anatole.chessel@polytechnique.edu}\\
}

\begin{document}
\maketitle

\begin{abstract}
  \textit{BioImageLoader} (BIL) is a python library that handles bioimage datasets for machine learning applications, easing simple workflows and enabling complex ones. BIL attempts to wrap the numerous and varied bioimages datasets in unified interfaces, to easily concatenate, perform image augmentation, and batch-load them. By acting at a per experimental dataset level, it enables both a high level of customization and a comparison across experiments. Here we present the library and show some application it enables, including retraining published deep learning architectures and evaluating their versatility in a leave-one-dataset-out fashion.
\end{abstract}


\section{Introduction}
\label{sec:intro}

Machine learning (ML) has taken biology by storm and bioimage informatics (BII) is one of the first discipline impacted. But core methodological advances and impressive application \cite{kraus2017automated, weigert2018content,stringer2021nm, greenwald2022whole} are the tip of the iceberg; the larger part of ML work involves a large number of careful numerical experiments comparing a variety of models, their hyperparameters, datasets or training regimen across losses and metrics. This implies specialized software to handle all those data, metadata and results. This operational side of ML, abbreviated as MLOps, has gathered a lot of attention across machine learning, with for example the libraries MLflow \url{https://mlflow.org} or Catalyst \url{https://catalyst-team.com} or start-ups Hugging Face \url{https://huggingface.co} or Weights \& Biases \url{https://wandb.ai}.

In BII specifically, one can cite ZeroCostDL4Mic \cite{vonchamier2021nc} which tries and provides a high level interface to train or study deep learning models. It implements an `MLOps for all' approach by using a free web application as well as free but limited resources provided by Google.
Other linked efforts includes the Bioimage Model Zoo \cite{ouyang2022}, a free, open, and community-driven hub on web for deep learning models for bioimages that helps developers to share and deploy their models and ImJoy \cite{ouyang2022}, which aims to integrate the whole MLOps through web interfaces. DeepImageJ \cite{gomez2021deepimagej} is an ImageJ plugin aiming at using models, in particular from the model zoo above, within ImageJ.

We propose \textit{BioImageLoader} (BIL), a python library to facilitate the handling of image datasets for ML workflows (Fig \ref{theFig}.A).
It introduces the experimental dataset as a unit, corresponding to the practical, daily use in biology of a set of images of similar sample, condition and imaging protocol.
By building individual wrapper around each such dataset, BIL make it easy to scale numerical experiments across many datasets, tailoring them to the specifics of each experimental datasets. In particular, it allows to assess the versatility of ML models by training them in a leave-one-dataset-out fashion. In the following we briefly present BIL and some applications it directly enables, including the retraining of common segmentation architecture on larger datasets. The code is open-source under BSD-3 license, and full documentation is available at \url{https://github.com/LaboratoryOpticsBiosciences/bioimageloader}, with the library being available to install through PyPI.

\section{Implementation}
\label{sec:impl}

\begin{figure}
  \centering
  \includegraphics[width=\textwidth]{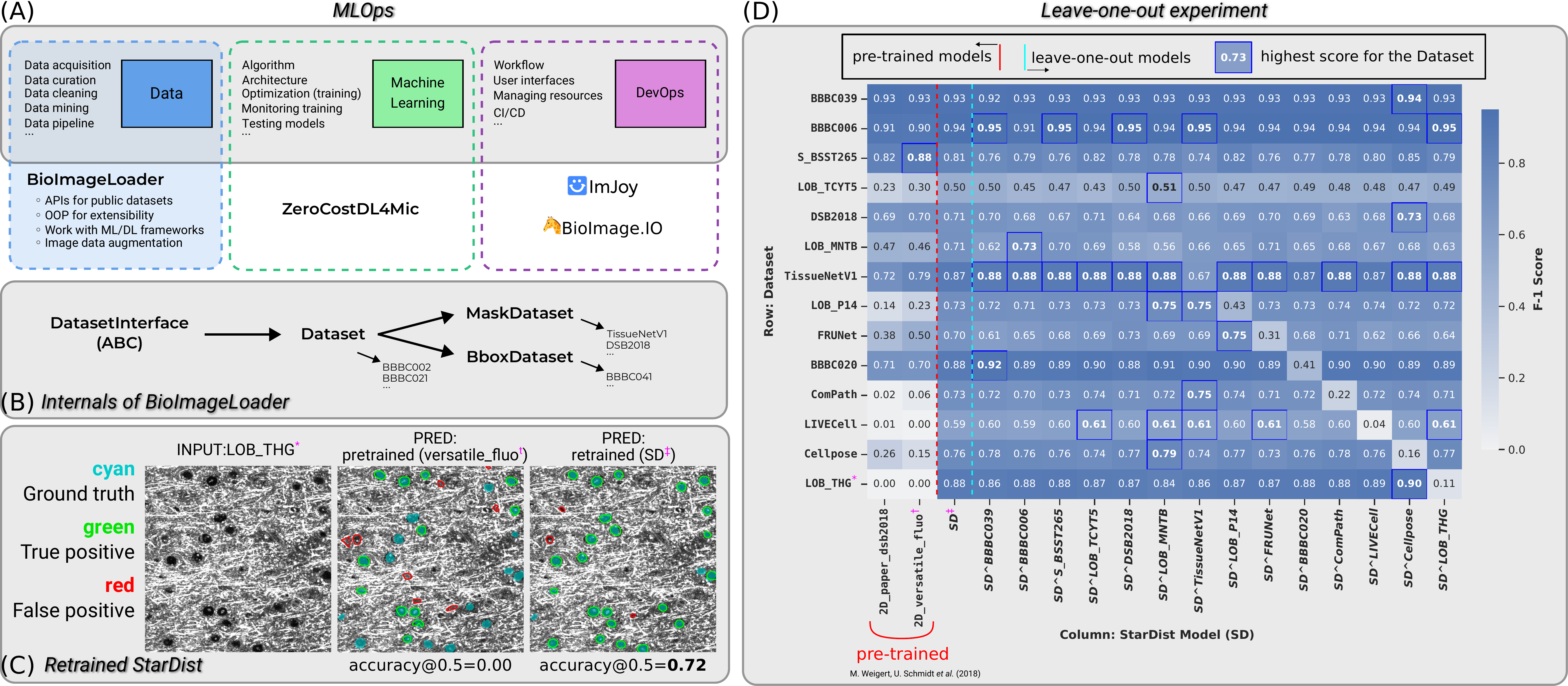}
  \caption{(\textbf{A}) MLOps refers to three main pillars: data, machine learning and DevOps. BioImageLoader (BIL) is a tool to handle data, which was missing. (\textbf{B}) BIL follows OOP (object-oriented programming) approach and makes use of ABC (abstract base class) to provide unified interfaces while keeping originality of each collection or dataset. (\textbf{C}) An example of a retrained StarDist architecture \cite{schmidt2018ac} over 14 datasets on one of four datasets we are releasing, called LOB\_THG (refer annotations ($\ast, \dagger, \ddagger$) to figure D on its right). (\textbf{D}) An ablation study in leave-one-out fashion that could be easily enabled by BIL. Column represents StarDist models that were either pretrained, retrained, or trained in leave-one-out fashion. Row represents datasets used to train or test models. Performance was measured by F-1 score.}
  \label{theFig}
\end{figure}

BIL is a Python programming library. It followed object-oriented programming (OOP) scheme using abstract base classes (ABCs). Each dataset interface is called a collection and is based on the same ABC which allows sharing common properties and methods while keeping originality with concrete classes (cf Fig \ref{theFig}.B). It prioritized compatibility and extensibility with popular ML libraries such as \texttt{PyTorch} and \texttt{TensorFlow} as well as a data augmentation library, in particular, \texttt{albumentations}. Separate configuration file can be used to manage settings and to help keep track of individual experiments with ease. For additional usability, it makes use of cache and batch loading for high performance.

Every collection implemented original structures of each dataset and intended ways to load data as well as its metadata. They expose APIs (application programming interfaces) and allows either shared operations or individual manipulation. Basic capabilities shared across datasets includes unified interface for data and annotation, easy concatenation, batching, train/test splitting, and data augmentation. Custom changes per datasets would include specific I/O tailored to the specific way each dataset files is organized, selecting channels, normalizing value ranges, \textit{etc}.

We have so far built interfaces for 28 open data datasets across many disciplines, sample and modalities, with new ones being easy to add. They include 17 datasets with annotations. We chose to not include the datasets themselves to avoid licensing issues, beyond the scope of this work, but they are all available online. In addition, we are releasing four additional annotated datasets from the Laboratory of Optics for Biosciences (LOB) with more challenging 2D segmentation problems from non-linear THG microscopy and multiphoton multicolor brainbow samples. They will be available on \url{https://zenodo.org}.

\section{Applications}
\label{sec:apl}

Example applications are developed at \url{https://github.com/sbinnee/nunet}. In particular, we can more easily look at image clustering across datasets to explore data similarity or retraining/fine-tuning of models on specific datasets. Specifically, the table in Fig. \ref{theFig}.D shows the retraining of StarDist architecture \cite{stringer2021nm}, one of the current state-of-the-art instance segmentation neural networks, with Fig \ref{theFig}.C showing an example on one of the new LOB dataset. Suggested model was trained across 14 datasets, either in a leave-one-dataset-out fashion, to assess the versatility of the models, or across all available datasets, to build a new generalist pretrained StarDist model; see \url{https://laboratoryopticsbiosciences.github.io/bioimageloader-docs/notebooks/train_models.html} for details. The pretrained model demonstrated in this example is made available to the community on Bioimage Model Zoo (\url{https://bioimage.io}) \cite{ouyang2022}.

\section{Conclusions}
\label{sec:cl}

Machine learning will revolutionize the way we do biology but only if running the many numerical experiments to build methods, scale them to larger dataset and/or tailor them to specific problems and workflow is not too cumbersome. We present \textit{BioImageLoader} a python library to help in handling bioimage datasets. A particular exciting perspective is scaling up current deep learning studies toward ever larger datasets, the main limitation keeping us away from proper generic models. Part of a larger MLOps effort, we believe community wide adoption and development of such open source tools will enable rapid growth in the application of ML to biology.

\section{Availability}
\label{sec:av}
BioImageLoader (BIL) is an open-source project. Source code is available at \url{https://github.com/LaboratoryOpticsBiosciences/bioimageloader}, and user manual at \url{https://laboratoryopticsbiosciences.github.io/bioimageloader-docs/}. The four new annotated datasets will be available on \url{https://zenodo.org}.

\bibliographystyle{unsrt}
\bibliography{main}  

\begin{thebibliography}{1}

\bibitem{kraus2017automated}
Oren~Z Kraus, Ben~T Grys, Jimmy Ba, Yolanda Chong, Brendan~J Frey, Charles
  Boone, and Brenda~J Andrews.
\newblock Automated analysis of high-content microscopy data with deep
  learning.
\newblock {\em Molecular systems biology}, 13(4):924, 2017.

\bibitem{weigert2018content}
Martin Weigert, Uwe Schmidt, Tobias Boothe, Andreas M{\"u}ller, Alexandr
  Dibrov, Akanksha Jain, Benjamin Wilhelm, Deborah Schmidt, Coleman Broaddus,
  Si{\^a}n Culley, et~al.
\newblock Content-aware image restoration: pushing the limits of fluorescence
  microscopy.
\newblock {\em Nature methods}, 15(12):1090--1097, 2018.

\bibitem{stringer2021nm}
Carsen Stringer, Tim Wang, Michalis Michaelos, and Marius Pachitariu.
\newblock Cellpose: A generalist algorithm for cellular segmentation.
\newblock {\em Nature Methods}, 18(1):100--106, January 2021.

\bibitem{greenwald2022whole}
Noah~F Greenwald, Geneva Miller, Erick Moen, Alex Kong, Adam Kagel, Thomas
  Dougherty, Christine~Camacho Fullaway, Brianna~J McIntosh, Ke~Xuan Leow,
  Morgan~Sarah Schwartz, et~al.
\newblock Whole-cell segmentation of tissue images with human-level performance
  using large-scale data annotation and deep learning.
\newblock {\em Nature biotechnology}, 40(4):555--565, 2022.

\bibitem{vonchamier2021nc}
Lucas {von Chamier}, Romain~F. Laine, Johanna Jukkala, Christoph Spahn, Daniel
  Krentzel, Elias Nehme, Martina Lerche, Sara {Hern{\'a}ndez-P{\'e}rez},
  Pieta~K. Mattila, Eleni Karinou, S{\'e}amus Holden, Ahmet~Can Solak,
  Alexander Krull, Tim-Oliver Buchholz, Martin~L. Jones, Lo{\"i}c~A. Royer,
  Christophe Leterrier, Yoav Shechtman, Florian Jug, Mike Heilemann, Guillaume
  Jacquemet, and Ricardo Henriques.
\newblock Democratising deep learning for microscopy with {{ZeroCostDL4Mic}}.
\newblock {\em Nature Communications}, 12(1):2276, April 2021.

\bibitem{ouyang2022}
Wei Ouyang, Fynn Beuttenmueller, Estibaliz {G{\'o}mez-de-Mariscal}, Constantin
  Pape, Tom Burke, Carlos {Garcia-L{\'o}pez-de-Haro}, Craig Russell, Luc{\'i}a
  {Moya-Sans}, Cristina {de-la-Torre-Guti{\'e}rrez}, Deborah Schmidt, Dominik
  Kutra, Maksim Novikov, Martin Weigert, Uwe Schmidt, Peter Bankhead, Guillaume
  Jacquemet, Daniel Sage, Ricardo Henriques, Arrate {Mu{\~n}oz-Barrutia}, Emma
  Lundberg, Florian Jug, and Anna Kreshuk.
\newblock {{BioImage Model Zoo}}: {{A Community-Driven Resource}} for
  {{Accessible Deep Learning}} in {{BioImage Analysis}}, June 2022.

\bibitem{gomez2021deepimagej}
Estibaliz G{\'o}mez-de Mariscal, Carlos Garc{\'\i}a-L{\'o}pez-de Haro, Wei
  Ouyang, Laur{\`e}ne Donati, Emma Lundberg, Michael Unser, Arrate
  Mu{\~n}oz-Barrutia, and Daniel Sage.
\newblock Deepimagej: A user-friendly environment to run deep learning models
  in imagej.
\newblock {\em Nature Methods}, 18(10):1192--1195, 2021.

\bibitem{schmidt2018ac}
Uwe Schmidt, Martin Weigert, Coleman Broaddus, and Gene Myers.
\newblock Cell {{Detection}} with {{Star-convex Polygons}}.
\newblock {\em arXiv:1806.03535 [cs]}, 11071:265--273, 2018.

\end{thebibliography}

\end{document}